\documentclass[
superscriptaddress,
reprint,
showpacs,
preprintnumbers,
nofootinbib,
nobibnotes,
amsmath,
amssymb, 
aps,
prd,
floatfix
]{revtex4-2}

\usepackage[utf8]{inputenc}
\usepackage[normalem]{ulem}
\usepackage{graphicx}
\usepackage{dcolumn}
\usepackage{bm}
\usepackage[colorlinks=true,allcolors=purple]{hyperref}
\usepackage{url}
\usepackage{enumerate}
\usepackage{array}
\newcolumntype{L}{>{\centering\arraybackslash}m{7cm}}

\usepackage{slashed,multirow,relsize,soul,feynmp-auto,tikz}
\usepackage{color}
\usepackage{mathrsfs} 
\usepackage{amsmath}
\usepackage{cancel}
\usepackage{cleveref}
\usepackage{bbold}
 \usepackage{mathrsfs}
\usepackage{braket}
\usepackage{physics}
\usepackage{multirow}
\usepackage{isotope}
\usepackage{slashed}
\usepackage{dcolumn}
\usepackage{latexsym}
\usepackage{verbatim}
\usepackage{float}
\usepackage{multirow}
\usepackage{xspace}

\newcommand{\refref}[1]{Ref.~\cite{#1}}

\newcounter{CommentCount}
\definecolor{MH}{rgb}{0.0,0.6,9}
\definecolor{palatinate}{rgb}{0.494, 0.192, 0.482}

\definecolor{DM}{rgb}{0.313,0.859,0.391}

\renewcommand{\phi}{\varphi}

\begin{document}

\preprint{\hfill FTPI-MINN-21-27}

\author{Matheus Hostert}
\email{mhostert@perimeterinstitute.ca}
\affiliation{School of Physics and Astronomy, University of Minnesota, Minneapolis, MN 55455, USA}
\affiliation{William I. Fine Theoretical Physics Institute, School of Physics and Astronomy, University of
Minnesota, Minneapolis, MN 55455, USA}
\affiliation{Perimeter Institute for Theoretical Physics, Waterloo, ON N2J 2W9, Canada}

\author{David McKeen}
\email{mckeen@triumf.ca}
\affiliation{TRIUMF, 4004 Wesbrook Mall, Vancouver, BC V6T 2A3, Canada}

\author{Maxim Pospelov}
\email{pospelov@umn.edu}
\affiliation{School of Physics and Astronomy, University of Minnesota, Minneapolis, MN 55455, USA}
\affiliation{William I. Fine Theoretical Physics Institute, School of Physics and Astronomy, University of
Minnesota, Minneapolis, MN 55455, USA}

\author{Nirmal Raj}
\email{nraj@iisc.ac.in}
\affiliation{Centre for High Energy Physics, Indian Institute of Science, C. V. Raman Avenue, Bengaluru 560012, India}

\title{Dark sectors in neutron-shining-through-a-wall and nuclear absorption signals}
\date{\today}
\begin{abstract}
We propose new searches for $n^\prime$, a dark baryon that can mix with the Standard Model neutron. 
We show that IsoDAR, a proposal to place an intense cyclotron near a large-volume neutrino detector deep underground, can look for $n\to n^\prime \to n$ transitions with much lower backgrounds than surface experiments.
This opportune neutron-shining-through-a-wall search would be possible without any modifications to the primary goals of the experiment and would provide the strongest laboratory constraints on the $n$-$n^\prime$ mixing for a wide range of mass splitting.
We also consider dark neutrons as dark matter and show that their nuclear absorption at deep-underground detectors such as SNO and Borexino places some of the strongest limits in parameter space.
Finally, we describe other $n^\prime$ signatures, such as neutrons shining through walls at spallation sources, reactors, and the disappearance of ultracold neutrons.
\end{abstract}

\maketitle

\section{Introduction} 

The existence of new baryons in a hidden sector has been a topic of great theoretical and phenomenological interest. One particularly interesting possibility is that of a dark neutron $n^\prime$, a new fundamental or composite dark particle that mixes with the Standard Model (SM) neutron. In addition to appearing in mirror sectors or brane world theories, $n^\prime$ is interesting on its own due to the potential impact on a number of observables. In cosmology and astrophysics, dark neutrons have been invoked to explain dark matter~\cite{McKeen:2015cuz,Karananas:2018goc,ClineCornellCosmo,utahcapture:2020bzz}, the baryon asymmetry of the Universe~\cite{McKeen:2015cuz,Aitken:2017wie,Elor:2018twp,Alonso-Alvarez:2019fym, Nelson:2019fln,Elahi:2021jia}, and help realize asymmetric inflation~\cite{Berezhiani:1995am,Babu:2021mjg}. They were also shown to modify the physics of cosmic rays, the cosmic microwave background, Big Bang nucleosynthesis, and neutron stars (NS)~\cite{GZK:Berezhiani:2005hv,Berezhiani:2006je,Berezhiani:2011da,McKeenNelsonReddyZhouNS,DBCosmoAstro:McKeen:2020oyr}.
In the laboratory, dark neutrons can appear in a variety of exotic particle physics processes~\cite{GZK:Berezhiani:2005hv,Klopf:2019afh,Elahi:2020urr,McKeen:2020zni,McKeen:2020vpf,FornalGrinsteinReview,Heeck:2020nbq,Strumia:2021ybk}. 
Among these are hydrogen decays, linked to an excess in XENON1T~\cite{XENON:2020rca,McKeen:2020vpf} (now superseded by XENONnT data~\cite{XENON:2022ltv}) and $n$-$n^\prime$ transitions that can explain the discrepancy between bottle~\cite{Serebrov:2004zf,Pichlmaier:2010zz,Steyerl:2012zz,Ezhov:2014tna,Arzumanov:2015tea,Pattie:2017vsj,Serebrov:2017bzo,UCNt:2021pcg} and beam~\cite{Byrne:1996zz,Nico:2004ie,Yue:2013qrc} measurements of the neutron lifetime.
This disagreement has been the subject of several new physics proposals.
One such class adds new exotic decay channels for the neutron~\cite{Fornal:2018eol,Cline:2018ami,Berezhiani:2018udo,Barducci:2018rlx}; however, these come to the cost of adding tension with recent data on the axial coupling $g_A$~\cite{Czarnecki:2018okw,Dubbers:2018kgh}.
Other proposals involve exotic $n-n^\prime$ transitions in the cold neutron beams~\cite{Berezhiani:2018eds}.

Motivated by the above, this paper explores the phenomenology of the dark neutrons $n^\prime$ in a general context. All that is assumed about $n^\prime$ is that it is a neutral state carrying unit baryon number that mixes with the SM neutron with an arbitrary mixing amplitude $\epsilon_{n n^\prime}$.
The low-energy two-state Hamiltonian of the $n$-$n^\prime$ system is
\begin{equation}
H = \begin{pmatrix} 
  m_n + \Delta E & \epsilon_{nn^\prime} \\
  \epsilon_{nn^\prime} & m_n + \delta m
 \end{pmatrix}~,
\label{eq:H}
\end{equation}
where $m_n$ is the neutron mass and 
$\Delta E$ is the energy contributed to the neutron by some matter-induced potential.
The parameter $\delta m \equiv m_{n^\prime} - m_n$ is the in-vacuum $n$-$n^\prime$ mass splitting that persists in the limit 
$\Delta E, \epsilon_{nn^\prime}\to 0$.
Facilities with ultracold neutrons (UCN)~\cite{PSI:Ban:2007tp,PSIlike:Serebrov:2009zz,PSI:Altarev:2009tg,Berezhiani:globalexpt:2017jkn,nEDM:2020ekj} and cold neutron beams~\cite{Broussard:2021eyr} have looked for $n \to n^\prime$ transitions between $n$ and $n^\prime$, with rates determined by the parameters in Eq.~\eqref{eq:H}.

We propose exploring a complementary neutron-shining-through-a-wall signature at an accelerator setup located underground.
Motivated by recent experimental activity in this area, we consider IsoDAR~\cite{Alonso:2021jxx,Alonso:2021kyu}, a proposal for a high-intensity cyclotron to be placed near a large volume detector underground, currently being considered for operation at Yemilab in Korea.
The IsoDAR setup would emerge as the most sensitive laboratory probe of dark neutrons to date, partly due to the copious number of neutrons produced by the accelerator, but mainly due to its underground location, where atmospheric backgrounds are minimal.
IsoDAR would provide a marked improvement over the strongest laboratory constraints set by STEREO~\cite{STEREO:Almazan:2021fvo}, cold neutron beams~\cite{Broussard:2021eyr}, and by UCNs at the nEDM experiment~\cite{nEDM:2020ekj}, all of which constitute surface experiments.

In recasting the UCN nEDM, we discuss how to interpret $n$-$n^\prime$ oscillation searches in terms of the more general model above, providing complementary coverage in a wider parameter space. We encourage collaborations to go beyond the assumption of a mirror symmetry lifted only by the magnetic or mirror magnetic field, and quote results in terms of the more general model in Eq.~\eqref{eq:H} (see \Cref{subsec:UCNdisap} for details).

For some parameters, the dark neutron is stable on cosmological time scales and could be the dark matter of the Universe, an attractive scenario wherein baryon number ensures the stability of both normal and dark matter~\cite{McKeen:2015cuz,Karananas:2018goc,ClineCornellCosmo,McKeenNelsonReddyZhouNS,utahcapture:2020bzz}.
We place strong limits on this scenario by using the measured absorption rate of single neutrons at the underground large-volume neutrino detectors SNO and Borexino.
These limits further complement the probes mentioned above.
All in all, our strategies probe new regions of parameter space not previously constrained by UCN searches and astrophysics.

This paper is laid out as follows.
In \Cref{sec:lims}, we describe our signals for both neutron-shining-through-a-wall setups and the absorption of dark neutron dark matter,
and estimate constraints and future sensitivities.
In \Cref{sec:other} we discuss alternative probes of dark neutrons that may also be of interest and sketch their sensitivities.
We then conclude and discuss our findings in \Cref{sec:concs}.

\begin{figure}[t]
    \centering
    \includegraphics[width=0.49\textwidth]{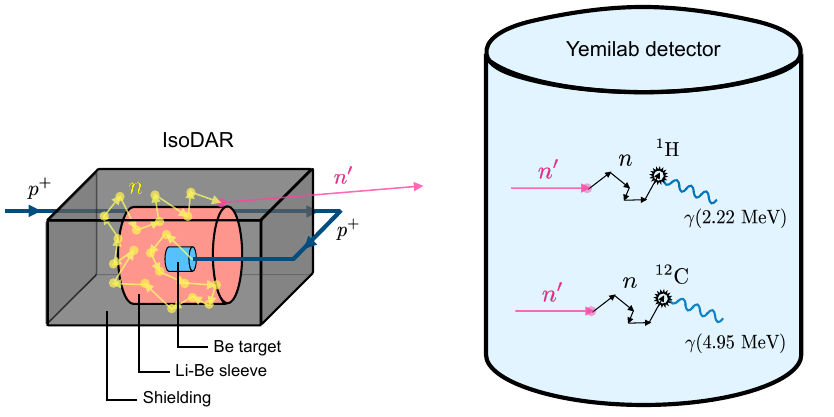}
    \caption{The neutron-shining-through-a-wall signature at IsoDAR and the detection processes in the Yemilab liquid-scintillator detector~\cite{Alonso:2021jxx,Alonso:2021kyu}.
    In the figure, we show the $\isotope[7]{Li}-\isotope[9]{Be}$-enriched sleeve enveloping the $\isotope[9]{Be}$ target. The sleeve is designed to increase the number of neutrons produced as well as their lifetime inside the sleeve~\cite{Bungau:2018spu}, therefore also maximizing the number of $n\to n^\prime$ conversions. On the right we show two detection possibilities: neutron capture on $\isotope{H}$ or $\isotope{C}$. Capture on Gd can also be considered, producing $\sim8$ MeV photons.\label{fig:IsoDARsetup}}
\end{figure}

\begin{figure*}[t]
    \centering
    \includegraphics[width=0.49\textwidth]{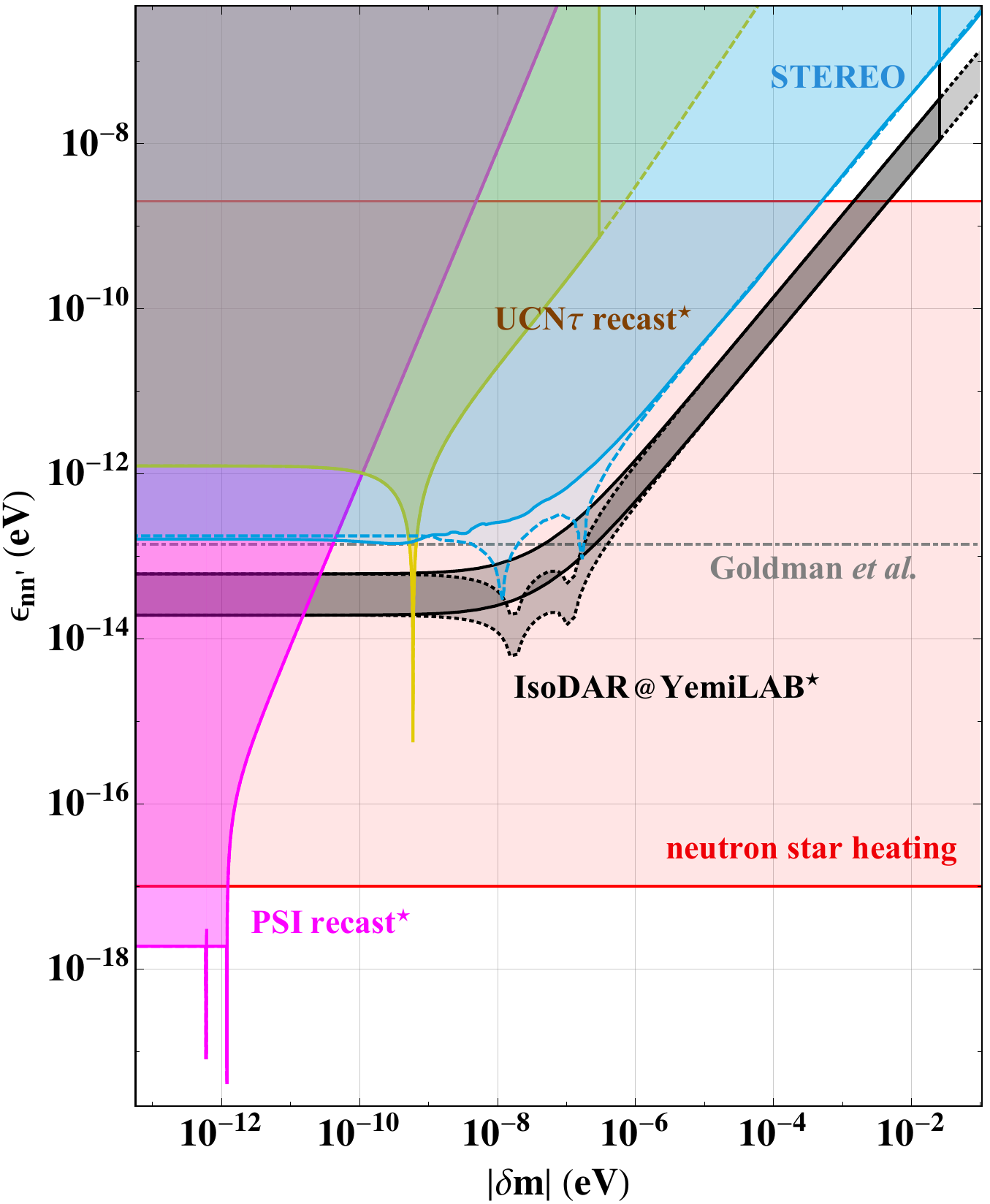}  \includegraphics[width=0.49\textwidth]{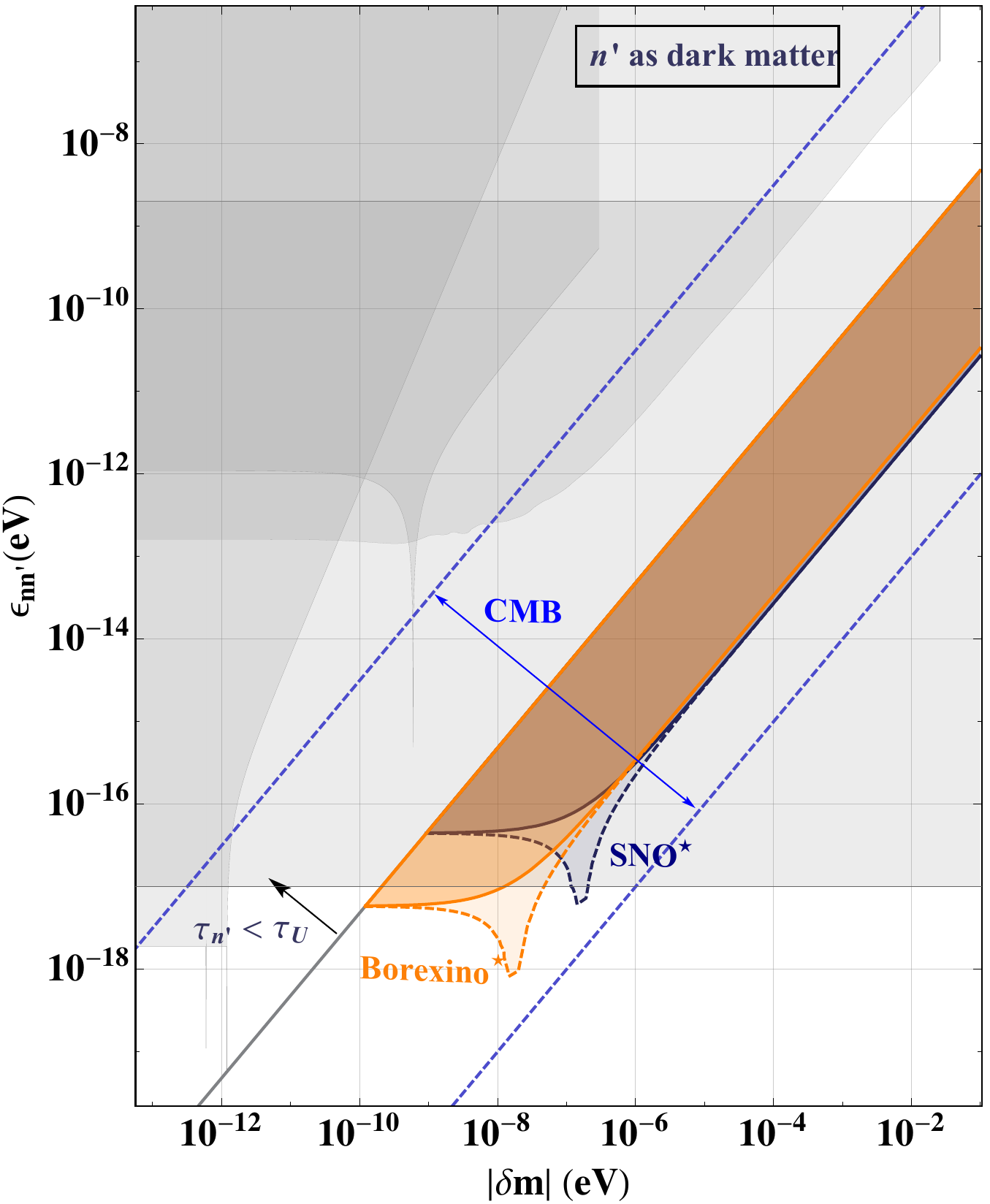} 
    \caption{
    Limits on the neutron-dark neutron transition amplitude $\epsilon_{nn'}$ as a function of the absolute mass splitting $|\delta m|$.
    The solid (dashed) curves correspond to $\delta m > 0$ ($\delta m < 0$), and labels marked with a star correspond to limits derived in this work.
    {\bf \em Left.} We show the 10 events/year sensitivity of the near-future IsoDAR experiment at YemiLAB as a black band. The top curve corresponds to the pessimistic scenario (capture on C only) while the bottom curve corresponds to the optimistic one (capture on H or $0.1\%$ Gd). The recent limits posed by the STEREO experiment are shown as cyan regions, which can be improved by IsoDAR by one to two orders of magnitude.
    These probes are complemented at small mass splittings by constraints from the non-observation of disappearance of ultracold neutrons (UCN) from their traps.
    In magenta we show our reinterpretation of a search at PSI~\cite{nEDM:2020ekj} and in yellow our limit derived from the neutron lifetime measurement at UCN$\tau$~\cite{UCNt:2021pcg}.
    Also shown are recently derived limits from overheating of neutron stars [red region] and a bound by Goldman {\em et al.} using pulsar rotation periods [gray dot-dashed line] (see also \cite{Goldman:2022brt} for interesting caveats).
    {\bf \em Right.} Limits for the case of the dark neutron constituting the Galactic dark matter [purple regions].
     The upper bound comes from the non-observation of extra neutrons in the measurement of the solar boron-8 neutrino flux at SNO and in a search for exotic decays of carbon at Borexino.
     The lower bound comes from demanding that the lifetime of the dark neutron exceeds the age of the universe.
     Also shown are limits from CMB observables on the decay $n^\prime \to p e \nu$. 
     The grey region in the background is the combination of limits shown on the left panel
     See text for further details.\label{fig:lim}}
\end{figure*}

\section{Signals and constraints}
\label{sec:lims}

In this section, we investigate neutron-shining-through-a-wall ($n \to n^\prime \to n$) probes and derive limits on $n^\prime$ as dark matter from neutron absorption signals, $n^\prime \to n$.
The resulting limits and sensitivities derived in our study are displayed in \Cref{fig:lim}. We compare them with existing limits from NS overheating (red region) derived in Ref.~\cite{Smoke&Mirrors:McKeen:2021jbh}, applicable for $\delta m$ up to $\mathcal{O}$(10) MeV as set by the NS nuclear potential. 
The gray dot-dashed line shows the upper bound on $\epsilon_{nn'}$ derived by Goldman {\em et al.}~\cite{Goldman:2019dbq} from the change in the rotation speed of NSs induced by the production of $n^\prime$ in their cores.
Also shown are limits from searches for $n \to n^\prime$ oscillations at the PSI UCN facility~\cite{nEDM:2020ekj} (magenta region) which we describe later.

We now describe the signatures in more detail. We start with the neutron-shining-through-a-wall rate. 
The net probability of $n \to n'$ conversion after time $t$~\cite{CowsikNussinov:1980np,wavepackettrap:Kerbikov:2008qs} is
\begin{equation}
P_{nn'}(t) = \frac{t}{t_f} \sin^2 2\theta \sin^2 \left[\frac{\sqrt{(\delta m - \Delta E)^2 + 4\epsilon_{nn'}^2}\, t_f}{2}\right]~,
\label{eq:probgen}
\end{equation}
where the in-medium mixing angle is given by $\tan 2 \theta = 2 \epsilon_{nn^\prime}/(\delta m - \Delta E)$ and
$t_f$ is the time-of-flight between collisions with the target material.
Note that the above expression is only valid when the neutron survives in the experiment for times longer than $t_f$. In the case of vacuum, this requires  $t_f < \tau_n$. In matter, this condition is more stringent and requires $t_f$ to be smaller than the average absorption time, which can be much smaller than $\tau_n$. In setting our limits, we assume that both conditions are satisfied. While neutrons are not efficiently absorbed in some materials, we note both conditions above are satisfied in the experimental setups considered here.
In addition, it neglects any phase-space suppression due to the mass difference of $n$ and $n^\prime$. We study this issue in detail in \Cref{app:phasespace}.

The pre-factor of $t/t_f$ can be intuitively interpreted as the number of collisions $N_{\rm coll}$, which gets multiplied by the oscillation probability to give the net conversion probability. The number of collisions can be estimated as $N_{\rm coll} = \Sigma_{\rm el}/\Sigma_{\rm abs}$, where $\Sigma_{\rm el (abs)}$ is the column density of the material for elastic scattering (absorption). Therefore, the conversion probability is maximized for materials with low absorption cross section, such as heavy water $\isotope{D}_2 \isotope{O}$.

The effective splitting $\Delta E$ could receive contributions from multiple sources: 
\begin{equation}
    \Delta E = {\bm \mu}_n \cdot \bm{B} + V_{\rm F} + ... \ ,
\end{equation}
where the first term is the Zeeman splitting induced by an external magnetic field, with $|{\bm \mu}_n | = 1.91 e/(2 m_p)$,
and the second term is the Fermi pseudopotential from neutron forward scattering in matter.
The latter is given by
\begin{equation}
   V_{\rm F} = \sum_i \frac{2\pi}{m_n} n_{\rm i} f_{{\rm scat},i}~, 
\end{equation}
where
$n_{\rm i}$ is the number density of nuclide $i$ in the scattering target material, and
$f_{\rm scat}$ is the neutron scattering length for a given target~\cite{fscatNIST}.

We now discuss the calculation of the signal rate for neutrons shining through a wall. At a distance $d$ away from a neutron point source, the flux of dark neutrons can be estimated as 
\begin{equation}
\Phi_{n'} = P_{nn'} \frac{R_n}{4\pi d^2},
\end{equation}
where $R_n$ is the neutron production rate and the $n \to n^\prime$ conversion probability is given in Eq.~\eqref{eq:probgen}.
The rate of detection of regenerated neutrons is then given by
\begin{equation}
    \Gamma^{\rm det}_{n^\prime\to n} = \sum_i N_{\rm i} \theta^2_{\rm det} \left( \sigma_{\rm i}^{\rm el}\varepsilon^{\rm el} + \sigma_{\rm i}^{\rm abs}\varepsilon^{\rm abs} \right) \Phi_{n'}
\end{equation}
where $i$ = \{C, H\} labels the target nuclide, 
$N_{\rm i}$ is the number of target nuclides in the detector, and
$\theta_{\rm det}$ is the mixing angle inside the detector material. The neutron cross section and efficiency of detection are denoted by $\sigma$ and $\varepsilon$, respectively.
In brackets we have two terms, the first corresponding to the conversion of $n^\prime\to n$ in an elastic scattering process followed by the capture of $n$, and the second corresponding to the conversion of $n^\prime\to n$ in the capture process. The first term often dominates since $\sigma^{\rm el} \gg \sigma^{\rm abs}$, unless the liquid scintillator (LS) is doped with $\isotope{Gd}$, whose capture cross section can be large enough to compensate for the typical $0.1\%$ concentration.

Throughout this work, we use 
\begin{equation}
\begin{aligned}
    &\sigma_{\rm H}^{\rm el} = 82~\text{b},\,\, \sigma_{\rm \isotope[7]{Li}}^{\rm el} = 1.4~\text{b}, \\
    &\sigma_{\rm \isotope[9]{Be}}^{\rm el} = 7.6~\text{b}, \,\,\sigma_{\rm \isotope[12]{C}}^{\rm el} = 5.6~\text{b},
\end{aligned}
\end{equation}
for thermal-neutron elastic scattering and 
\begin{equation}
\begin{aligned}
    \sigma_{\rm \isotope[12]{C}}^{\rm abs}(E_{\gamma}=4.95~{\rm MeV}) &= 3.5\times 10^{-3}~\text{b},\\
    \sigma_{\rm \isotope{H}}^{\rm abs}(E_{\gamma}=2.22~{\rm MeV}) &= 3.3\times 10^{-1}~\text{b},\\
    \sigma_{\rm \isotope[155]{Gd}}^{\rm abs}(E_{\gamma}=8.54~{\rm MeV}) &= 6.1\times 10^{4}~\text{b},\\
    \sigma_{\rm \isotope[157]{Gd}}^{\rm abs}(E_{\gamma}=7.94~{\rm MeV}) &= 2.5\times 10^{5}~\text{b},\\
    \sigma_{\rm \isotope[7]{Li}}^{\rm abs} = 4.5\times 10^{-2}~\text{b},\,\, \sigma_{\rm \isotope[9]{Be}}^{\rm abs} &= 7.6\times 10^{-3}~\text{b},
\end{aligned}
\end{equation}
for thermal-neutron capture cross section~\cite{neutroncrosssecs}. 
For simplicity, we assume that the efficiency of detection in elastic and absorption conversions are the same, though the former is expected to be slightly lower due to the possibility of losing the converted neutron.
In practice we set $\varepsilon^{\rm el} = \varepsilon^{\rm abs} = 0.3$~\cite{STEREO:Almazan:2021fvo}.
We also assume that $N_{\rm C} = N_{\rm H}$, as is the case for benzene and mineral oil.

\subsection{Neutrons-shining-through-a-wall at IsoDAR}

IsoDAR is a proposal to place a high-power cyclotron near a large-volume neutrino detector~\cite{Bungau:2012ys}. The proton beam creates neutrons in large numbers, which in turn create $\isotope[8]{Li}$, that finally decay at rest to electron antineutrinos. This source can be used for a high-precision and high-intensity program for neutrino physics, including $\bar \nu_e$ disappearance searches as a test of light sterile neutrinos~\cite{Diaz:2019fwt,Dasgupta:2021ies} as well as electroweak precision measurements~\cite{Conrad:2013sqa}.
While the cyclotron concept has been extensively studied~\cite{Bungau:2018spu,2019NatRP...1..533A}, 
the detector choice and siting options are to be determined.
Previous proposals considered placing IsoDAR near the KamLAND detector~\cite{Abs:2015tbh,Alonso:2017fci} as well as pairing it with water-based liquid scintillator detectors such as THEIA~\cite{OrebiGann:2015gus}. 
Recently, however, the possibility of siting IsoDAR at the Yemilab underground laboratories in South Korea has been under active consideration~\cite{Alonso:2021jxx,Alonso:2021kyu}. Motivated by this renewed interest, we consider the sensitivity of IsoDAR@Yemilab to dark neutrons. When appropriate, we also rescale our estimates for IsoDAR@KamLAND, noting that both locations benefit from a low rate of cosmogenic backgrounds.

The neutron fluence at IsoDAR is smaller than that of modern neutron spallation facilities and nuclear reactors by factors of $\mathcal{O}(10)$ and  $\mathcal{O}(100)$, respectively.
Nevertheless, it provides an advantageous setup, given that it is located underground. The large overburden provided by the deep location of the proposed IsoDAR sites suppresses the rate of cosmogenic neutrons to negligible levels as compared to surface experiments. In addition, IsoDAR's proximity to a kt-scale detector maximizes its sensitivity to new rare phenomena and allows for greater fiducialization to shield external backgrounds. Neutrons from the source can also be efficiently shielded with a few meters of material.

We now turn to the experimental details.
The IsoDAR cyclotron delivers a 10~mA current of 60 MeV protons on a $\isotope[9]{Be}$ target, corresponding to $6.25\times 10^{16}$ protons per second~\cite{Alonso:2017fci}. 
The neutron production efficiency is approximately $10\%$ yielding an intense source of $R_n \simeq $ 6.25 $\times 10^{15}$ neutrons/second.
In the current design~\cite{Alonso:2021jxx}, the beam target is surrounded by a sleeve with 30\% to 70\% mass ratio of high-purity $\isotope[7]{Li}$ and $\isotope[9]{Be}$, respectively, which upon absorbing neutrons, produces $\isotope[8]{Li}$ in large numbers.
The $\isotope[8]{Li}$ isotopes subsequently undergo beta-decay at rest to produce electron-antineutrinos $\bar\nu_e$ with an endpoint energy of $E_\nu \sim 15$~MeV. 

The sleeve is designed to maximize the number of neutrons by reducing absorption on $\isotope[6]{Li}$ impurities and increasing the production of secondary neutrons~\cite{Bungau:2018spu}. Fortuitously, this also serves to maximize the rate of $n^\prime$ production; see \Cref{fig:IsoDARsetup} for a schematic of our setup.
In Eq.~\eqref{eq:probgen} we replace $t/t_f \to N_{\rm coll}$, and approximate the number of collisions as $N_{\rm coll} = (\Sigma^{\rm el}/\Sigma^{\rm abs})_{\rm Li-Be} \approx 300$.
We note, however, that in a full experimental analysis, it would also be important to include the finite size of the sleeve and the collisions of neutrons in the shielding that surrounds it.

For the detector, we assume a $2.5$~kt liquid-scintillator (LS) detector with a fiducial mass of $1.16$~kt, with its center situated $17$ m away from the target~\cite{Alonso:2021kyu}. For LS, neutrons can be captured on $\isotope{H}$, emitting $2.2$~MeV gammas, as well as  on $\isotope{C}$, emitting $4.95$~MeV gammas. The former is $f_{\rm H}/f_{\rm C} \times \sigma_{n{\rm H}}/\sigma_{n{\rm C}} \simeq 94$ times more common, where $f_i$ is the fraction of $\isotope{H}$ and $\isotope{C}$ in the scintillator, which we assume to be the same. Nevertheless backgrounds for the latter are expected to be much smaller.
We also note that a water-based liquid scintillator (WbLS) detector~\cite{Alonso:2014fwf,Land:2020oiz} is under consideration at Yemilab~\cite{Seo:2019dpr}. This may present a more challenging situation for a single neutron measurement due to the larger energy resolution and smaller $\isotope{C}$ concentration.

In all cases, in particular, for a WbLS detector, addition of $\isotope{Gd}$ to the detector volume would be greatly advantageous. Capture on $\isotope{Gd}$ would dominate over that of hydrogen or carbon, and produce $7.9$~MeV and $8.54$~MeV photons that are much harder to mistake for environmental backgrounds. In fact, for a typical $0.1\%$ Gd concentration, the ratio between capture on Gd and H is approximately
\begin{equation}\label{eq:GdtoH}
   \sum_{i=155,157} 0.1\% \times  N_{{\rm Gd{\text -}}i}\, \sigma^{\rm abs}_{{\rm Gd{\text -}}i} \simeq N_{\rm H}\, \sigma^{\rm abs}_{\isotope{H}}.
\end{equation}
We exploit this correspondence below.

\subsubsection{Backgrounds}

We now discuss potential backgrounds to a search for reappearing neutrons inside the Yemilab detector.
A background study for single-hit $\overline{\nu}_e e^-\to \overline{\nu}_e e^-$ events at IsoDAR@Yemilab was performed in~\refref{Alonso:2021kyu} under the assumption of a LS detector.
In contrast to neutrino-electron scattering, our neutrons-shining-through-a-wall signature consists of a single mono-energetic photon from neutron capture, narrowing the energy region of interest. 
In addition, if the Yemilab detector technology allows for angular reconstruction (see, for example, progress in angular reconstruction in LS recently reported by Borexino~\cite{Borexino:2021qgg,Borexino:2021dmz}, as well as studies for WbLS in \refref{Land:2020oiz}), solar and beam-related neutrino-electron scattering backgrounds could be further reduced since the neutron capture emits gammas isotropically. 

\paragraph{Neutrino-induced backgrounds} 
The first background to consider comes from mis-identified inverse beta decay (IBD) events. 
When the positron or the neutron is undetected, IBD can fake a single-hit event.
Reference~\cite{Alonso:2021kyu} finds that over 5 years about $0.25\%$ of the  $1.67\times 10^{6}$ IBD events at IsoDAR@Yemilab would appear as a single hit.
This estimate considers only the detection of a single positron with a missing neutron. In the energy range of interest for capture on $\isotope{H}$, it is negligible, while for capture on $\isotope[12]{C}$, it constitutes a total of $2.8$ events. We note, however, that the alternative case of IBD events with a missing positron and a single neutron could also present a potentially more serious background to our search. If the scattering takes place inside the active or veto volume of the detector, it could be efficiently identified due to the two $511$~keV photons from the positron annihilation inside the detector. However, if the positron is produced in a blind region, the neutron can leak inside the fiducial volume and mimic our signal. An evaluation of this background is left to future work in the hope that it be carried out with more sophisticated simulations. In case the rates are large, one should consider smaller fiducial volumes.

In addition to IBD, neutral-current (NC) neutrino-nucleus interactions can also produce a single neutron inside the detector. For $\isotope[12]{C}$, the neutrino energy threshold for such a process is $E_\nu > 18.7$~MeV, which is larger than the energies of both $\isotope[8]{Li}$ and solar $\isotope[8]{B}$ decays. However, $\nu + \isotope[13]{C} \to \nu + \isotope[12]{C}(\text{g.s.})$ has a lower threshold of $E_\nu > 4.95$~MeV~\cite{Arafune:1988hx,Fukugita:1989wv,Suzuki:2012aa,Suzuki:2019cra}. This process is expected to take place in the Yemilab LS, albeit at a small rate due to the natural abundance of $\isotope[13]{C}$ of $\sim1\%$. Using the cross section in \refref{Suzuki:2019cra}, we calculate the ratio between the IBD and $\overline{\nu}_e$-induced single-neutron-knockout NC interactions to be $6.7\times 10^{-6}$. For the $2.26$~kt fiducial volume of the IBD analysis, \refref{Alonso:2021kyu} quotes $1.67\times 10^{6}$ IBD events in 5 years of running, which corresponds to $5.7$ free neutrons in the $1.16$~kt fiducial volume of the single-hit analysis. An additional $2.6$ free neutrons are expected from $\isotope[8]{B}$ solar neutrinos. Similar processes also produce the excited state $\isotope[12]{C}(2+)$, but at a lower rate and effectively tagged due to the presence of $E_{\gamma}\sim 4.44$~MeV de-excitation photons.
Finally, the LS may also contain some concentration of $\isotope[2]{H}$, which has a much lower threshold for NC dissociation. Taking, for example, the concentration reported by KamLAND~\cite{Tolich:2005gy}, $10^{-4}$, and the neutrino-deuteron cross section from Ref.~\cite{Nakamura:2000vp}, we find a total of $12$ and $3.1$ events in 5 years from $\isotope[8]{Li}$ antineutrinos and $\isotope[8]{B}$ neutrinos, respectively.
Although an irreducible background, the total $23$ NC interactions are sub-dominant with respect to environmental backgrounds discussed below and are unlikely to produce signals of neutron capture on C. 

While a signal in Ref.~\cite{Alonso:2021kyu}, $\overline{\nu}_e-e$ interactions represent a background for us. This component has a shape that is well-known and a normalization that would be constrained by \emph{in-situ} measurements. Using a Gaussian energy resolution of $6.4\%/\sqrt{E}$, we find a total of $16$ events within two standard deviations of the C capture energy, $4.95$~MeV. For capture on H, we find 63 events.

\paragraph{Environmental sources} 
Another source of backgrounds comes from radioactivity, spontaneous fission, and spallation in and around the detector.
Most backgrounds are smoothly distributed in visible energy and therefore can be constrained outside the peak energy for neutron capture on $\isotope{H}$, $\isotope{C}$, or $\isotope{Gd}$. 
From Fig.~5 of \refref{Alonso:2021kyu}, we can directly read off the backgrounds to a single-hit $n$ capture on carbon to be $16$ events in 5 years of operation. Therefore, provided that IBD events with a missing positron are sufficiently rare, neutron capture on $\isotope{C}$ would constitute a low-background search. 

For $2.2$~MeV energies no information available. It is fair to assume, however, that this rate is much larger, especially due to low-energy environmental gammas. For instance, the backgrounds at KamLAND increased by two to three orders of magnitude below $E_{\rm vis} < 3$~MeV, mostly due to radioactivity inside the detector~\cite{KamLAND:2009zwo,Grant:2012nht}. Therefore, a detailed evaluation of the radiopurity of the detector is necessary to obtain a realistic prediction for backgrounds in the low-energy region. We proceed, however, noting that by exploiting the characteristic waveform of the signal, Borexino was able to achieve very low neutron rates at $2.2$~MeV, with less than 57 events in 485 days for 533~t of LS~\cite{Borexino:PEPV:2009mcw}. 

In summary, we find two possibilities for a LS detector: neutron capture on $\isotope{C}$ with $E_\gamma = 4.95$~MeV and an estimated background of $32$~events, and neutron capture on $\isotope{H}$ with backgrounds that are expected to be larger, mostly driven by environmental factors. A more detailed evaluation of radiogenic backgrounds at around $2.2$~MeV would help confirm if a search for neutron capture on $\isotope{H}$ is at all competitive. We conclude emphasizing that, by virtue of \Cref{eq:GdtoH}, it is possible to interpret our constraints in the H case as an estimate for a Gd-doped detector, where backgrounds would be much smaller.

\subsubsection{Sensitivity}

Having discussed some of the channels of interest, we now discuss the resulting sensitivity of IsoDAR@Yemilab to the $n$-$n^\prime$ mixing angle. We draw two curves: a pessimistic sensitivity based on neutron capture on $\isotope{C}$ only, and an optimistic sensitivity based on neutron capture on $\isotope{H}$. In both cases we require $\Gamma^{\rm det}_{n}$ = 10 events/yr. While this is realistic for the former scenario, it may seem too strong of a requirement for capture on H. This is justified, however, if the detector is doped with $\isotope{Gd}$ $0.1$~\% concentration. In this case, the capture rate is as large as the one in Hydrogen, and backgrounds can be kept at or under the levels of those of capture on C.

The resulting sensitivities in $\epsilon_{nn'}$ as a function of $\delta m$ are shown in \Cref{fig:lim} for both $\delta m > 0$ and $\delta m < 0$. The band corresponds to the difference between optimistic and pessimistic sensitivities.
We cut off the $\delta m > 0$ curves at 0.03 eV as that is the kinetic energy down to which neutrons are refrigerated by scatters in the shielding; above this mass splitting, $n \to n^\prime$ conversions are kinematically forbidden for thermal neutrons.
The dips in the curves correspond to resonances at the Fermi pseudopotential, with $V_{\rm F}$ = 116~neV ($17$~neV) for the Li-Be sleeve (liquid scintillator).  
Up to background considerations, the corresponding sensitivities for IsoDAR@KamLAND would not be all that different; with a $0.897$~kt fiducial mass of liquid-scintillator and a detector center located $16.1$~m from the target, the event rates are $\sim 1.15$ times smaller.
As the rate $\propto \epsilon_{nn'}^4$, this results in a reach in $\epsilon_{nn'}$ only 1.03 times weaker than shown for IsoDAR@Yemilab in \Cref{fig:lim}.

Although producing significantly more neutrons per second than IsoDAR, STEREO's search was limited by backgrounds from cosmogenic neutrons.
Thanks to IsoDAR's placement deep underground, this background can be effectively mitigated and one can improve the sensitivity to $\epsilon_{nn'}$ by 1--2 orders of magnitude, as seen in \Cref{fig:lim}. 

\subsection{UCN disappearance}
\label{subsec:UCNdisap}

The fact that significant losses of ultracold neutrons from their traps have not been observed can be exploited to place limits on $n \to n^\prime$ conversions since the feebly interacting $n^\prime$ can escape through the walls.
References~\cite{PSI:Ban:2007tp,Serebrov:2008her,PSIlike:Serebrov:2009zz,PSI:Altarev:2009tg,Berezhiani:globalexpt:2017jkn,nEDM:2020ekj,UCNt:2021pcg} report results on searches for neutron disappearance as a function of a mirror magnetic field $B'$ under the assumption of $\delta m = 0$. 
As discussed, we can reinterpret these for the case of a non-zero $\delta m$ to set upper limits on the plane of $\epsilon_{nn'}$ versus $\delta m$.
Here we describe our procedure for reinterpreting these limits.

First, we note that typical UCN facilities confine neutrons to below a kinetic energy KE$_{\rm max}$ = 300 neV~\cite{UCNPSIReview}, corresponding to the optical potential of the trap material.
This corresponds to a time-of-flight $t_f \lesssim \mathcal{O}(10^{-2}~{\rm s})$ for meter-scale traps~\cite{nEDM:2020ekj}.
The UCNs receive an additional effective mass of $|\Delta E| \simeq 10^{-12}$~eV $\simeq 1500$~Hz sourced by an externally applied field.

Thus for $|\Delta E - \delta m| \gg t^{-1}_f$ we are in the ``fast oscillation" regime, so that in Eq.~\eqref{eq:probgen} we can set the second squared sine to 1/2, giving
\begin{equation}
P_{nn'} \to \frac{t_s}{t_f} \frac{2\epsilon^2_{nn'}}{4\epsilon^2_{nn'}+(\delta m - \Delta E)^2}~,
\label{eq:probUCNfast}
\end{equation}
where $t_s$ is the storage time in UCN traps.
Conversely, for $|\Delta E - \delta m| \ll t^{-1}_f$ (in the limit $\epsilon_{nn'} \ll t^{-1}_f$, which will shortly be seen to be true), we are in the ``slow oscillation" regime and we get
\begin{equation}
P_{nn'} \to \epsilon^2_{nn'} t_s  t_f~.
\label{eq:probUCNslow}
\end{equation}
Therefore, we get a finite probability on resonance ($\delta m = \Delta E$) that depends on the parametric splitting between the mass eigenstates, $\epsilon_{nn'}$.

Having discussed the two limiting cases of $n-n^\prime$ oscillations in UCN experiments, we proceed to reinterpret two separate measurements as constraints on $\epsilon_{nn'}$.

\subsubsection{The ratio method -- $E_B$}

Motivated by the suggestions in Ref.~\cite{Berezhiani:2009ldq}, the neutron electron dipole moment (nEDM) collaboration at PSI has performed a search for an anomalous disappearance of unpolarized UCNs~\cite{nEDM:2020ekj}.
The search was based on the count rate of neutrons 
in setups with and without magnetic fields applied.
By comparing these rates under different magnetic fields, they placed limits on models with mirror magnetic fields.
These limits exclude various positive results reported in the literature~\cite{PSI:Ban:2007tp,Serebrov:2008her,Berezhiani:2012rq,Berezhiani:2017jkn}.

The collaboration also set a 95\% C.L. limit $\tau_{n n^\prime}= 1/\epsilon_{n n^\prime} > 352~\text{s}$ on the disappearance of $n$ under the assumption of $\Delta E = 0$ (no mirror magnetic fields), which holds for the case $\delta m = 0$.
This translates to a limit on the transition amplitude for the two different applied magnetic fields ($B = 10~\mu$T and $B = 20~\mu$T): 
\begin{equation}\label{eq:nEDMlimit}
    \epsilon^{95\%, \delta m = 0}_{nn'} < 1.87 \times 10^{-18}~{\rm eV}~.
\end{equation}

The constraint above is derived using the ratio method, where the following observable is defined,
\begin{equation}
    1 + E_B = \frac{2 n_{0}}{n_B + n_{-B}} = \frac{2 e^{-m_s P^{B=0}_{nn^\prime}} }{ e^{-m_s P^{B}_{nn^\prime}} +  e^{-m_s P^{-B}_{nn^\prime}}},
\end{equation}
with $n_{i}$ the number of neutrons counted after a storage time $t_s$ in a setup with an applied magnetic field $i$. 
The last equality shows the relevant expression for $B^\prime = 0$, where $m_s = t_s/\langle t_f\rangle$ is the average number of bounces off the wall. 
Here we assume $t_s = 380$~s and $\langle t_f\rangle = 0.085$~s.
The searches use three $B$-field configurations: $B=0$, $\pm 10$~$\mu$T, $\pm 20$~$\mu$T. 
The last configuration provides the strongest limits, so we proceed to assume $|B| = 20$ $\mu$T.

In the limit of small transition probabilities $m_s P_{nn^\prime}$, which will be the case for the parameter space of interest, we can rewrite $E_B$ as
\begin{equation}\label{eq:EBequation}
    E_B \simeq - m_s \left(P^{B=0}_{nn^\prime} - \frac{ P^{B}_{nn^\prime} + P^{-B}_{nn^\prime} }{2}\right),
\end{equation}
which for $\delta m \to 0$ is the relevant expression for the constraint in Eq.~\eqref{eq:nEDMlimit}. 
By virtue of the large time-of-flight of the neutrons in the experiment ($\Delta E \ll t_f$) and assuming $\langle t_f \rangle \simeq t_f = 0.085$~s, in the slow-oscillation regime we have
\begin{equation}
    E_B \simeq - m_s t_f^2 \epsilon_{nn^\prime}^2,
\end{equation}
which together with Eq.~\eqref{eq:nEDMlimit} gives the limit on the observable $E_B^{\rm nEDM} < 2.6\times 10^{-4}$ at $95\%$ C.L. 
With Eq.~\eqref{eq:EBequation}, this allows us to set the constraints across the relevant plane of $\epsilon_{nn^\prime}$ versus $\delta m$ using the oscillation probability in Eq.~\eqref{eq:probgen}.
The resulting limits are shown as the magenta region in \Cref{fig:lim} at $95\%$~C.L.; the two peaks correspond to the resonances at the Zeeman energies. 
For large mass splittings, we get
\begin{equation}
    \epsilon_{nn^\prime} < 2.6 \times 10^{-11} \left( \frac{\delta m }{1 \text{ neV}}\right)^2\left( \frac{ 1.2 \text{ peV} }{\Delta E}\right)\left(\frac{ E_B}{2.6 \times 10^{-4}}\right),
\end{equation}


We now describe the scaling behavior of the the observable $E_B$ at large mass splittings. 
For $\epsilon,\Delta E \ll \delta m$, the first  non-zero term in the expansion of Eq.~\eqref{eq:EBequation} gives
\begin{equation}
    E_B \simeq - 6 m_s \frac{\epsilon^2_{n n^\prime} |\Delta E|^2}{\delta m^4}.
\end{equation}
Therefore, the limits on $\epsilon$ at large mass splittings should scale as $\delta m^2$ and have a much steeper slope than other limits shown in this paper, which at large $\delta m$ are effectively limits on $\theta = \epsilon_{nn^\prime}/\delta m$.

\subsubsection{The lifetime method -- $\tau_n$}

Another measurement that is sensitive to additional $n$ disappearance in UCN experiments is that of the neutron lifetime, using the bottle method. 
In this case, the observable is simply the rate of disappearing neutrons, which is translated as a measurement of the neutron lifetime. 
In the presence of $n-n^\prime$ oscillations, however, the experimental observable is instead
\begin{equation}
    \left(\tau_{n}^{\rm exp}\right)^{-1} = \Gamma_{n \, {\rm decay}}^{\rm SM} + \Gamma_{nn^\prime},
\end{equation}
where, on the right-hand side, the first term is just the neutron decay rate in the SM and the second the rate of $n\to n^\prime$ transition in the experiment. 
We can then translate the measurement of the lifetime into constraints on $\epsilon$ and $\delta m$. 
For instance, at very large mass splittings ($\Delta E, t_f^{-1} \ll \delta m$), the transition rate is simply $\Gamma_{n n^\prime} = 2 \epsilon_{nn^\prime}^2/t_f \delta m^2$, and the latest and most precise measurement of the neutron lifetime with UCN by the UCN$\tau$~\cite{UCNt:2021pcg}, $\tau_n^{\text{UCN}\tau} = (877.75 \pm 0.36)$~s, together with the SM prediction $\tau_n^{\text{SM}} = (878.7 \pm 0.6)$~s, provides a $95\%$~C.L. limit on the mixing parameter,
\begin{equation}
    \epsilon^{95\%\text{ C.L.}} < 6.6 \times 10^{-13}\text{ eV} \left(\frac{\delta m}{1\text{ neV}} \right) \left( \frac{t_f}{0.1\text{ s}} \right)^{\frac{1}{2}}.
\end{equation}
When $\Delta E$ is comparable to the vacuum mass splitting, the transition may be resonantly enhanced, and the above formula no longer holds. 
Similarly, when $|\delta m| \ll \Delta E$, the transition probability is no longer dependent on $\delta m$, and the constraint on $\epsilon_{n n^\prime}$ saturates. 
Our full limit is shown as the yellow line in Fig.~\ref{fig:lim}, and it shows that this saturation happens in a region already excluded by the ratio method, discussed above.
Nevertheless, as alluded to above, the lifetime method dominates the constraint at large mass splittings due to its softer dependence on $\delta m$.
The solid (dashed) curve depicts $\delta m > 0 \ (\delta m < 0)$.
The $\delta m > 0$ curve is cut off at $\delta m =$ KE$_{\rm max}$ as $n \to n^\prime$ conversions are kinematically forbidden beyond that.
No such cutoff exists for the $\delta m < 0$ case, where downscatters are kinematically always allowed.

\renewcommand{\arraystretch}{1.5}
\begin{table*}[t]
    \centering
    \begin{tabular}{|c|c|c|c|}
    \hline
      Type & Experiment & $\Delta E$ (eV) & Limit on $\theta_{\rm vac} \equiv \theta({\Delta E = 0})$ \\
        \hline
        \hline
        \multirow{5}{*}{shining-through-a-wall} 
                %
                %
                & { MARS @ SNS }~\cite{COHERENT:2021qbu} &  {$\sim 10^{-8}$} & {$6.3 \times 10^{-5}$} \\\cline{2-4}
                & { MURMUR}~\cite{Stasser:2020jct} & ${\sim 10^{-8}-10^{-7}}$ & { $1.4 \times 10^{-5}$} \\\cline{2-4}
                & { STEREO}~\cite{STEREO:Almazan:2021fvo} & ${\sim 10^{-8}-10^{-7}}$ & { $3.9 \times 10^{-6}$} \\\cline{2-4}
                & {IBR-2}~\cite{Shabalin:2015hga} &  {$\sim 10^{-8}$} & {$9.3 \times 10^{-5}$} \\\cline{2-4}
                & \textbf{ IsoDAR@YemiLAB}~\cite{Alonso:2021jxx,Alonso:2021kyu} & \bm{ $\sim 10^{-8}$} & \begin{tabular}{c} \bm{ $1.3 \times 10^{-6}$} (capture on $\isotope{H}$)\\ \bm{$ 4.0 \times 10^{-7}$} (capture on $\isotope{C}$) \end{tabular} \\\cline{2-4}
        \hline 
        \hline
        ultra cold neutron disappearance &   {nEDM}~\cite{nEDM:2020ekj} & {$\sim \pm 10^{-12}$} & {$1.8 \times 10^{-6}$} \\
        \hline 
        \hline
        \multirow{2}{*}{resonant neutron regeneration} & {SNS \emph{Broussard et al}}~\cite{Broussard:2021eyr} & -- & $[1,10] \times 10^{-5}$~\footnote{Valid in the range $\delta m = [50,400]$~neV.} \\
        & \textbf{HFIR}~\cite{HFIRWallShine:Broussard:2017yev,Broussard:2019tgw} & -- & \bm{$\epsilon_{nn^\prime} < 4.4\times10^{-17}$}~\textbf{eV}~\footnote{No information available for the $\delta m \gg \Delta E$ regime.} \\
        \hline 
        \hline
        \multirow{2}{*}{atmospheric neutrons} &   {SNO}~\cite{SNO:2003bmh} & \multirow{3}{*}{{$10^{-12}$}}   &  {$5.9 \times 10^{-4}$}\\
              & {Borexino}~\cite{Borexino:PEPV:2009mcw} &  &  {$6.0 \times 10^{-4}$} \\
        \hline
                \hline
        dark neutron dark matter & {SNO}~\cite{SNO:2003bmh} & {$1.7 \times 10^{-7}$} & {$2.7 \times 10^{-10}$}\\
          (nuclear absorption)     & {Borexino}~\cite{Borexino:PEPV:2009mcw}  &  {$1.7 \times 10^{-8}$} & {$3.3 \times 10^{-10}$} \\ 
        \hline 
    \end{tabular}
    \caption{Summary of the exclusion limits imposed by experiments discussed in this work. Bold entries represent future projections.
    Experimental limits are quoted at the $95\%$~C.L. level, except for Borexino, which are $90\%$~C.L. The IsoDAR projection corresponds to $10$~events/year.
    Here, $\Delta E$ is the effective mass contribution to the neutron from some potential sourced by the scattering medium or ambient magnetic fields; see Eq.~\eqref{eq:H}.
    Limits are quoted on the in-vacuum mixing angle ($\theta_{\rm vac} = \epsilon_{nn'}/\delta m$ for $\delta m \gg \Delta E$). 
    \label{tab:otherprobes}
    }
    
\end{table*}

\subsection{Dark neutrons as dark matter: absorption signals}
\label{subsec:dndm}

Dark neutrons were likely in chemical equilibrium with SM states in the early universe and can be expected to have primordial abundances comparable to that of standard baryons~\cite{DBCosmoAstro:McKeen:2020oyr}.
If cosmologically long-lived, they could then constitute the dark matter of the universe.
We call this scenario dark neutron dark matter (DNDM).

We can probe this scenario by detection of excess neutrons via $n^\prime \to n$ conversion of the DNDM flux.
Reference~\cite{utahcapture:2020bzz} estimated future sensitivities at neutrino and dark matter experiments using nuclear capture signals.
Here we show that extensive limits on DNDM may be already placed by considering past measurements at SNO and Borexino.

Taking the local DM density $\rho_\odot = 0.3$~GeV/cm$^3$ and average speed $v_\chi$ = 270 km/s~\cite{recommend:Baxter:2021pqo}, the mean DM flux $\Phi_\chi = (\rho_\odot/m_{n'}) v_\chi  = 8.6 \times 10^6$/cm$^2$/s. 
For keV kinetic energies, the neutron-proton scattering cross section $\sigma_{np}$ = 4 b~\cite{NNOnline}, thus the rate-per-proton with which neutrons would be detected from the DNDM flux is 
\begin{equation}
R_{n'/p} = \Phi_\chi \times \theta^2 \sigma_{np} =  \theta^2  \times 3.5 \times 10^{-17}~{\rm s}^{-1}~, 
\label{eq:DNDMrateperprot}
\end{equation}
We compare this rate with the measured rate of single neutrons in underground neutrino detectors. 

\paragraph{Single neutrons at SNO} We first consider the total number of neutrons produced by $^8$B solar neutrinos in $\nu D \to \nu n p$, as measured by SNO~\cite{SNO:2003bmh}. In particular, we use the fact that the $\nu$ flux measured by SNO, $\simeq (5.21 \pm 0.67) \times 10^6$/cm$^2$/s, is in agreement with predictions.
Taking the NC cross sections at the relevant energies from Ref.~\cite{Butler:2000zp}, and the $^8$B $\nu_x$ fluxes from Ref.~\cite{Billard:2013qya}, we get the rate per deuteron, (equivalent to the number of neutrons produced per deuteron) as
\begin{equation}
    R_{\nu/D} = 2.2 \times 10^{-36}~{\rm s}^{-1}~.
\end{equation}
We then demand that not more than 25\% (approximately a 95\% C.L. constraint) of these SNO neutrons are sourced by our regeneration mechanism, i.e. $R_{n'/p} \leq 0.25  R_{\nu/D}$, using $V_{\rm F} = 166$~neV for heavy water when computing the in-medium mixing angle.

\paragraph{Single neutrons at Borexino} 
Another experiment we consider is Borexino.
In particular, a search for ``Pauli Exclusion Principle violation" in $\isotope[12]{C}\to \isotope[11]{\widetilde C} + n$ transitions was performed by counting the 2.2 MeV photons produced via by neutron capture on $\isotope{H}$~\cite{Borexino:PEPV:2009mcw}.
Using a fiducial mass of 100 tonnes, a $90\%$~C.L. of $<57$ signal events was obtained in this channel.
This can be compared to Eq.~\eqref{eq:DNDMrateperprot} to set limits, where now $V_{\rm F} = 17$~neV for pseudocumene LS. 

Both the SNO and Borexino measurements result in the upper limits on $\epsilon_{nn'}$ versus $\delta m$ shown in \Cref{fig:lim} right panel.
Also shown in the gray background is the combined region excluded by other probes as displayed on the left panel.
The solid (dashed) curve corresponds to $\delta m > 0$ ($\delta m < 0$).
The lower limits (i.e., the ceiling of the exclusion region) are set by requiring that dark neutrons do not decay via the beta channel $n^\prime \to p e \nu$ within the age of the universe:
\begin{equation}
\tau_{n'} \geq  \tau_{\rm U} = 13.7~{\rm Gyr}~,
\label{eq:DMceilcriterion}
\end{equation}
with the decay rate\footnote{For $\delta m > 0$ the decay channel $n^\prime \to n \gamma$ is also open. However, this is subdominant to $n^\prime \to p e \nu$ for $\delta m<3~\rm eV$.} given by~\cite{DBCosmoAstro:McKeen:2020oyr} 
\begin{equation}
\tau_{n'}^{-1} =  \frac{1}{12 \ {\rm Gyr}} \bigg(\frac{\epsilon_{nn'}/\delta m}{5 \times 10^{-8}}\bigg)^2,
\end{equation}
where we have assumed that the $n$-$n^\prime$ mass splitting is small compared to the $Q$ value of neutron beta decay of $782~\rm keV$.

We have also displayed in \Cref{fig:lim} right panel the region that would be excluded by cosmic microwave background (CMB) observations of the reionization history, impacted by the injection of electromagnetic energy from the final state electron in $n^\prime$ decay following the recombination epoch.
This limit applies for $n^\prime$ lifetimes in the range $10^{12}~{\rm s} \lesssim \tau_{n'} \lesssim 10^{26}~{\rm s}$~\cite{Slatyer1211,ClinePoorCMB}; see Ref.~\cite{DBCosmoAstro:McKeen:2020oyr} for further details.

We see that in the DNDM scenario we are able to probe parameter space that is complementary to the regions constrained in the previous two sub-sections.
In particular, we can limit (a) $\epsilon_{nn'}$ up to three orders of magnitude smaller than STEREO and IsoDAR, due to the detection rate being proportional now only to two powers of the mixing angle as opposed to four, and 
(b) $\delta m \gg$ 0.03~eV as, unlike in the case of searches initiated by $n \to n'$ conversions, keV dark neutrons are not kinematically limited by in-medium thermalization of neutrons. 

We also note that the DM can be comprised of anti-$n^\prime$ particles. 
In that case, their scattering on nuclei would lead to the appearance of anti-neutrons, which subsequently annihilate with neutrons and deposit $\sim$ 2 GeV energy in detectors. 
The signature of such DM will be very similar to the di-nucleon decay signal tightly constrained by the Super-Kamiokande collaboration~\cite{Super-Kamiokande:2020bov}. 
Ref.~\cite{antinprimeDM:Keung:2019wpw} determined that $\sim 10^6$ signal events would be registered in the multi-pion channel post kinematic cuts for a benchmark $\theta = 6 \times 10^{-11}$.
About 20 background events are expected in this region, thus this translates to a limit of $\theta \lesssim 4 \times 10^{-12}$~eV, which is more than an order of magnitude stronger than the SNO limit on $n^\prime$ DM (see Table~\ref{tab:otherprobes}).
This is not surprising given the spectacular signature of dinucleon decay at Super-K resulting in greater energy deposition and fewer backgrounds than nuclear capture at SNO.
Moreover, the CMB limits will also be correspondingly stronger, as the energy injection will be $O(10^3)$ times larger compared to the beta decay of $n^\prime$.

\section{Other probes}
\label{sec:other}

In this section, we discuss a few more probes of dark neutrons.
As we shall see, these are not as sensitive as the probes described in the previous section, but they still provide a significant amount of neutrons that may be leveraged with future experimental progress. In addition, they may also be relevant for dark baryon models that are more complicated than the one considered here.
In Table~\ref{tab:otherprobes} we enumerate these strategies alongside those discussed above and, to facilitate comparison, show their reaches in vacuum mixing angle, $\theta_{\rm vac}=\theta({\Delta E = 0})\simeq \epsilon_{nn^\prime}/\delta m$. 

\subsection{Surface neutron sources} 

\subsubsection{Reactors}

Recently, new limits on $n$-$n^\prime$ mixing were placed by the STEREO experiment~\cite{STEREO:Almazan:2021fvo} at the Institut Laue-Langevin (ILL) research reactor in France. This neutron-shining-through-a-wall search was originally proposed in \refref{Sarrazin:2015sua} and improved on previous limits by the MURMUR experiment~\cite{Sarrazin:2016bsw,Stasser:2020jct}, located at BR2 nuclear reactor in Belgium.
STEREO searched for neutrons sourced by the ILL reactor converting to dark neutrons via thermalizing scatters in the surrounding heavy water, and subsequently reappearing inside the liquid scintillator detector modules about 10 m away. The search is limited by atmospheric backgrounds, which can be subtracted by a measurement of the neutron rates during the reactor-off periods.

We note that Ref.~\cite{STEREO:Almazan:2021fvo} shows limits for up to $\delta m = 10^4$~eV.
However,  most neutrons at the STEREO source should thermalize down to room temperatures in the D$_2$O shielding, analogous to the thermalization of IsoDAR neutrons in the surrounding sleeve and shielding described above.
This would result in an analogous cutoff in $\delta m$ on the limits; we have drawn a vertical cyan line at $\delta m = 0.03$~eV to roughly indicate what we believe is the boundary of validity of STEREO's limits. The STEREO and MURMUR limits in the vacuum case are shown in \Cref{tab:otherprobes}.

\subsubsection{ORNL Spallation Neutron Source}

At the Spallation Neutron Source (SNS) at the Oak Ridge National Laboratory (ORNL), neutrons are produced by a $1$~GeV proton beam impinging on a $\isotope{Hg}$ target. The beam current is $I = 1$~mA with a repetition rate of $60$~Hz, corresponding to $1.04\times10^{14}$ protons per pulse. The estimated neutron rate is around $1.5\times10^{17}$ neutrons per second assuming 25 neutrons are produced for every proton.
These neutrons can serve as a $n^\prime$ source for the neutrino and neutron detectors used by the COHERENT collaboration~\cite{COHERENT:2017ipa,COHERENT:2020iec}.
In particular, MARS~\cite{COHERENT:2021qbu}, a neutron detector, was located at a distance of $19.5$~m away from the target and was used to understand the flux of fast neutrons at the location of the neutrino alley, where coherent neutrino-nucleus scattering experiments are located. 

The neutrino alley is located in a basement with a depth of $8$~m.w.e., which provides even less shielding against cosmic ray backgrounds than STEREO's location. This source, however, benefits from the pulsed nature of the beam, allowing for the further reduction of cosmic backgrounds by looking for signals in coincidence with the narrow beam pulses.

 The measurement in MARS was performed using a double-hit signature of an elastic neutron scattering on hydrogen followed by a delayed capture on $\isotope{Gd}$. The flux of prompt neutrons, defined as $t<2\mu{\rm s}$ from the beginning of the beam pulse, was measured for neutrons in the energy range of $3.5$~MeV to a few tens of MeV. The final measurement was quoted in terms of a neutron flux, $\Phi_n = 1.20 \pm 0.56$~neutrons/m$^2$/MWh. This quantity can be used to place a limit on neutrons-shining-through-a-wall by demanding that the rate of $n\to n^\prime\to n$ conversion do not exceed the observed rate. As an approximation, we neglect multiple scattering at the production and assume all neutrons produced to be within the relevant energy range used in the measurement. The resulting limit on the vacuum mixing angle $\theta_{\rm vac}$ is then 
\begin{equation}
\theta_{\rm vac} < 6.3\times 10^{-5},
\end{equation}
at $95\%$ C.L. assuming an integrated power of $7000$~GWh per year.

It is also possible to derive limits using delayed thermal neutrons. This would benefit from the increased number of neutron collisions around the target and from the efficacy of the shielding against the thermal visible neutrons. However, the cosmogenic backgrounds would increase, and the rate of observable double-hit events would be more suppressed.

\subsubsection{J-PARC Spallation Neutron Source}

The J-PARC Spallation Neutron Source (JSNS) is the second most intense neutron source in the world. Neutros are obtained from a $3$~GeV proton beam from the Rapid Cycle Synchrotron (RCS) at J-PARC that strike a $\isotope{Hg}$ target. The beam has $1$~MW power and comes in pulses of $8\times10^{13}$ protons/$\mu$s at a 25~Hz frequency. Extrapolating the neutron multiplicity due to the spallation processes up to $E_p = 3$~GeV energies, we find a rate of $\lesssim 50$ neutrons per proton. Overall, we estimate that JSNS produces approximately $1.5\times10^{17}$ neutrons per second.

The J-PARC Sterile Neutrino Search at JSNS (JSNS$^2$) experiment~\cite{Ajimura:2017fld} is currently operational and expects to collect several tens of thousands IBD events in the best-fit regions of the sterile-neutrino interpretation of the LSND results in the course of 3 years of operation. The detector~\cite{JSNS2:2021hyk} is located 24~m away from the Hg target (see also future plans for a second detector~\cite{Ajimura:2020qni}), and consists of a cylindrical volume containing a gamma-catcher and a veto volume on the outermost layers with $31$~t of LS (no Gd), and a neutrino target with 17~tons of Gd-loaded LS in the innermost volume. The IBD neutrons are captured by Gd after $\sim 30\,\mu$s, yielding an $\sim 8$~MeV photon, to be compared with the $200\,\mu$s-delayed photons with $\sim 2.2$~MeV energy from capture on H. The former is a preferred signal as the background of beam-related gammas are expected to be too severe up to energies of $2.6$~MeV~\cite{Harada:2015gla}. 

Single hit events, both prompt and delayed, are copious in the detector and are produced by cosmic rays as well as beam-related photons. The rate for a $25$~t detector was found to be $\sim 5.2\times 10^{5}$ events/9~$\mu$s, quoted for the duration of the ``prompt" signal region. This rate assumes a PID rejection larger than 100 to differentiate photons from neutrons, yielding more than $10^{7}$ events in the prompt window. Although there are techniques that can be explored to reduce the number of backgrounds, such as using the pulsed nature of the beam to reject beam-related backgrounds and reducing the detector volume to shield against fast neutrons, cosmic-induced neutrons provide a continuous source of backgrounds that cannot be avoided. A subtraction technique similar to the one employed by STEREO~\cite{STEREO:Almazan:2021fvo} could be explored. We encourage the collaboration to pursue further studies in this direction.

In addition to JSNS and SNS, the European Spallation Source will be well poised to make progress in this field given that its neutron fluence is expected to surpass that of both SNS and JSNS~\cite{Addazi:2020nlz}. 

\subsubsection{IBR-2 Neutron Pulse Reactor}

The long-running IBR-2 reactor at Dubna is the world's most intense source of pulsed neutrons\footnote{despite SNS' claim in the Guinness Book of World Records~\cite{Guinness}.}.
Sourced by 92 kg of PuO$_2$ of density 11.5 g/cm$^3$, its {\em in situ} flux is $10^{16}$ n/cm$^2$/s~\cite{AKSENOV1991438}, from which we get its intensity as $4\times 10^{16}$ n/s.
The total neutron flux over an energy range of $10^{-6}$-10~MeV was measured at a distance of 0.9~m as 0.58 n/cm$^2$/s~\cite{Shabalin:2015hga}.
Requiring this flux to be below that from $n\to n^\prime \to n$, we obtain a bound the vacuum mixing angle
\begin{equation}
\theta_{\rm vac} < 9.3\times 10^{-5}~.
\end{equation}

\subsection{Atmospheric neutrons shining through the ground}

Neutrons produced in the collisions of cosmic rays with atmospheric nuclei undergo several nuclear scatters before thermalizing down to non-relativistic speeds.
Operating in the interaction basis, these scatters serve as measurements of the interaction eigenstate during which $n \to n^\prime$ conversions can occur.
For small mixing angles, the $n^\prime$ flux produced is unlikely to scatter further, could go through the Earth's solid rock, and arrive at an underground detector, where it could regenerate as neutrons that may be detected. 
This process is similar in some respects to the one described in \Cref{subsec:dndm}, with two important differences: the event rate now goes as {\em four} powers of the mixing versus two, due to two extra powers in the rate of $n^\prime$ production, and the average $n^\prime$ energy is expected to be $\mathcal{O}$(eV), much lower than the keV energy of DNDM.

We will estimate the sensitivity of this probe at SNO and Borexino as in \Cref{subsec:dndm}.
To that end, we first take the atmospheric neutron flux from Ref.~\cite{10.1667/RR0610.1}, which we denote by $\Phi_n^{\rm atm}$. 
Next, in Eq.~\eqref{eq:probgen} we set $t/t_f$ (the number of elastic scatters at the top of the atmosphere before the neutron gets absorbed) to 100, roughly the ratio of scattering vs absorption cross sections of the average nuclide in the atmosphere.
We then make a simplification by recognizing that the Zeeman splitting due to the Earth's magnetic field, $\Delta E_{B_\oplus} = 3 \times 10^{-12}$~eV is much greater than the Fermi pseudopotential of air at the top of the atmosphere ($\Delta E_{\rm air} = 10^{-17}$~eV), where the neutron flux is highest.
Thus we set $\Delta E = \Delta E_{B_\oplus}$ in Eq.~\eqref{eq:probgen}.

The dark neutron flux is now $P_{nn'} \Phi_n^{\rm atm}$.
The neutron appearance rate at SNO and Borexino can now be computed as done in \Cref{subsec:dndm}, from which we obtain the limit $\theta_{\rm vac} \lesssim 6 \times 10^{-4}$ for $\delta m \gg \Delta E$.

\subsection{Resonant cold neutron regeneration}

In more controlled environments, like in cold neutron beams, a neutron-shinning-through-a-wall experiment can be carried out with the help of magnetic fields. By adjusting the spatial configuration of the magnetic field, one can arrange for $n\to n^\prime$ as well as $n^\prime \to n$ transitions to occur as neutrons and dark neutron beam passes through a series of resonances. 
Depending on the geometry, magnetic fields, and wavelength of the neutrons, this can drive neutrons to shine through walls. Most recently, this was explored at the SNS with a cold neutron beam passing through a $4.6$~T magnetic field~\cite{Broussard:2021eyr}. 
Due to the resonant nature of the signal, the constraints obtained are valid only within a specific range of $\delta m$. The range of the cold neutron limits are quoted in \Cref{tab:otherprobes}, applicable only to the $\delta m $ range of $20$~neV to $400$~neV.

A similar experiment can also be carried out at ORNL using the High Flux Isotope Reactor (HFIR), which produces even more neutrons than the spallation source. The original proposals in Refs.~\cite{HFIRWallShine:Broussard:2017yev,Broussard:2019tgw} quote a best sensitivity of $\epsilon_{nn'} > 4.4 \times 10^{-17}$~eV in the degenerate region, $\delta m \ll \Delta E$. We refrain from translating this sensitivity to $\theta_{\rm vac}$ since this would depend on the specifics of the experimental apparatus, but note that it should improve upon the SNS limits above due to the larger neutron intensity. 

\section{Discussion}
\label{sec:concs}

We have shown that the near-future IsoDAR@Yemilab setup is highly suitable for a neutron-shining-through-a-wall search for dark neutrons resulting in the sensitivities in \Cref{fig:lim}.
These sensitivities complement limits from NS heating and UCN disappearance.
While the NS limits are stronger at low $\delta m$, it is desirable to probe this parameter space in laboratory conditions as NS temperature measurements come with specific assumptions on nuclear astrophysics and astronomical uncertainties.
The NS heating band in \Cref{fig:lim} is limited from below by the rate at which $n \to n^\prime$ conversions overheat neutron stars; 
In addition, as recently pointed out in Ref.~\cite{Goldman:2022brt}, in well-motivated mirror sectors with kinetic mixing between the mirror and SM photon, there are new mechanisms for heat dispersion inside the NS and the limits shown in \Cref{fig:lim} are weakened.
UCN experiments are uniquely poised to test this region for small $\delta m$ due to their small $\Delta E$ arising from magnetic fields, and disappearance rates that scale only as two powers of $\epsilon_{nn'}$.
The NS heating band is also limited from above by the suppression of $n \to n^\prime$ conversion within the lifetime of the youngest NS observed due to saturation of the $n^\prime$ Fermi sea~\cite{Smoke&Mirrors:McKeen:2021jbh}; IsoDAR is already seen to probe higher $\epsilon_{nn'}$ for $\delta m > 10^{-3}$~eV.
IsoDAR is also seen to outperform UCN disappearance searches across a vast range of $\delta m$, simply because the population of UCNs in traps cannot compete with the high intensity of neutrons produced in accelerator/cyclotron setups.

One of the main goals of the IsoDAR proposal is the study of sterile neutrinos, which can be viewed as an example of a broader perspective of dark sector physics \cite{Agrawal:2021dbo}. 
Application of the IsoDAR experiment to the search of dark neutrons represents another  way this experiment can contribute to the studies of dark sectors. Other applications,  that are yet to be studied in the IsoDAR setup, may include the production of exotic unstable particles via nuclear reactions \cite{Izaguirre:2014cza} and production of light dark matter with its subsequent interaction in a neutrino detector \cite{Izaguirre:2015pva}.

Finally, we comment on the implications of our results on ultraviolet physics.
A natural UV completion of our setup with sub-eV mass splittings is a mirror sector of the SM giving rise to a dark neutron $n^\prime$ mixing with the SM neutron via the dimension-9 operator
\begin{align}\label{eq:UVconnect}
{\cal L}&\supset\frac{1}{\Lambda^5}uddu^\prime d^\prime d^\prime+{\rm h.c.}\nonumber
\\
&\to\frac{\left(4\pi f_\pi^3\right)^2}{\Lambda^5}nn^\prime+{\rm h.c.}
\\
&=10^{-10}~{\rm eV}\left(\frac{\rm TeV}{\Lambda}\right)^5nn^\prime+{\rm h.c.},\nonumber
\end{align}
where we have used naive dimensional analysis in moving to the effective Lagrangian at the hadronic level in agreement with lattice computations~\cite{Buchoff:2015qwa,Rinaldi:2019thf}. This higher dimensional operator arises from renormalizable couplings involving new states---for instance, a color triplet $R$-parity-violating squark (or anti-diquark) $\phi$ and a SM singlet $\chi$ with couplings to quarks and mirror quarks of the form $d^{(\prime)}\chi\phi^\ast$ and $u^{(\prime)}d^{(\prime)}\phi$ (see, e.g.,~\cite{Arnold:2012sd,Aitken:2017wie,BerezhianiDKnH:2018udo} for possible UV completions). Necessarily, some of these new states, like $\phi$, carry QCD charge and can be produced at hadron colliders like the LHC. As we can see in Eq.~(\ref{eq:UVconnect}), putting these new states at the TeV scale, the rough LHC limit on new QCD-charged particles, can easily lead to  a mixing amplitude $\epsilon_{nn'}$ of about $10^{-10}~\rm eV$. Obtaining larger mixing amplitudes that are above the neutron star cooling limits is possible if the SM-singlet states integrated out to get (\ref{eq:UVconnect}) are pushed down to the GeV scale, i.e. $m_\chi\ll m_\phi$, increasing $\epsilon_{nn'}$ by $\sim 10^3$ or more. Thus much of the parameter space we propose to probe can arise in plausible UV completions leading to the effective Hamiltonian in~(\ref{eq:H}) that we study.

\section*{Acknowledgments}
We gratefully acknowledge fruitful discussion with
Adriana Bangau,
Janet Conrad,
Shmuel Nussinov, 
Joshua Spitz,
and
Sandra Zavatarelli. 
We thank Bernhard Lauss for correcting our scaling of UCN bounds in a previous version of this manuscript,
and the anonymous referee for helpful suggestions including an estimate of the IBAR-2 sensitivity.
The work of D.\,M. (N.\,R.) is (was) supported by the Natural Sciences and Engineering Research Council of Canada. 
TRIUMF receives federal funding via a contribution agreement with the National Research Council Canada.
M.P. is supported in part by U.S. Department of Energy Grant No. desc0011842.
The research of M.H. was supported in part by Perimeter Institute for Theoretical Physics. 
Research at Perimeter Institute is supported by the Government of Canada through the Department of Innovation, Science and Economic Development and by the Province of Ontario through the Ministry of Research, Innovation and Science.

\appendix 
\onecolumngrid 

\section{Transition probabilities}
\label{app:phasespace}

In this appendix we derive the $n$-$n^\prime$ transition probabilities for a simplified potential.
We discuss the probability scaling with the $n$-$n^\prime$ mass splitting $\delta m$, and the limiting cases of UCNs and fast neutrons.

\subsection{Two States in a Bottle}

We start with neutrons inside a stationary, magnetic, and spherical ``bottle", either magnetic or mechanical. 
The time-independent Hamiltonian of the system in the interaction basis, $(n, n^\prime)^T$, is
\begin{equation}
H = 
\left(\frac{p^2}{2m_n} + m_n\right) \times \mathbb{1} 
+
\begin{pmatrix} 
  \Delta E & \epsilon_{nn^\prime} \\
  \epsilon_{nn^\prime} & \delta m 
\end{pmatrix}
+
\begin{pmatrix} 
    V(\vec{x}) & 0 \\
    0 & 0
\end{pmatrix}.
\label{eq:Happ}
\end{equation}
The interaction between the neutron $n$ and the reflective walls is parameterized by the potential energy $V_0$, approximated as a sharp transition at a radius $r=R$,
\begin{equation}
    V(\vec{x}) = V(r) = V_0 \Theta(r - R).
\end{equation}
In principle, an additional contribution to the self-energy of the two neutrons can arise from residual magnetic fields in the ordinary or mirror worlds, parameterized by $\Delta E$.
The rotation to the mass basis can be performed with the mixing angle
\begin{equation}
    \tan 2\theta(r)  = \frac{2 \epsilon_{nn^\prime}}{\delta m - \Delta E - V(r)},
\end{equation}
where we keep the radial dependence of the potential explicit and define the mixing angle inside the bottle as
$\tan 2\theta_0 \equiv 2\epsilon_{nn^\prime}/(\delta m - \Delta E)$.
The vacuum mass basis, $V(r) = 0$, is a convenient one, and the Hamiltonian becomes
\begin{align}\label{eq:Hm_full}
H_{\rm m} &= \left(\frac{p^2}{2m_n} + m_n + \Delta E\right) \times \mathbb{1} +
\frac{\delta m -\Delta E}{\cos{2\theta_0}} \begin{pmatrix} 
     - \sin^2{\theta_0} & 0 \\
    0 &  \cos^2{\theta_0}
\end{pmatrix}    
+
\frac{V(r)}{2}\begin{pmatrix} 
    1+\cos{2 \theta_0} & -\sin{2\theta_0} \\
    - \sin{2\theta_0} & 1-\cos{2\theta_0}
\end{pmatrix}.
\end{align}
In the limit $\theta_0\to 0$, we are left with two independent states, one of which is interacting. 
On the resonance, $\Delta E = \delta m$ (i.e., $\theta_0 \to \pi/4$), we find two degenerate states that interact will interact with the bottle walls with the same strength.

\subsection{Neutron-Mirror-Neutron Transitions} 

We are mainly interested in the transition between an initial mostly-active neutron inside the bottle to an escaping mostly-sterile neutron.
In particle collisions, the transition of a neutron to a more massive final state is endothermic, and ought to be suppressed; it will eventually shut off due to energy conservation.
On the other hand, if it transitions to a much lighter most-sterile state, the transition is exothermic and is unsuppressed as it is energetically favourable. 
Nevertheless, UCNs in a trapping device are not undergoing single-particle collisions.
Their de Broglie wavelegnth is far greater than the interatomic distance, and it interacts with a macroscopic object, namely, the walls of the bottle.

To understand the dependence on $\delta m$, we will neglect the residual energy splitting, $\Delta E \to 0$.
We start by expanding the Hamitonian in \Cref{eq:Hm_full} in the small mixing angle $\theta_0$. 
Separating the interaction Hamiltonian in powers of $\theta_0$,
\begin{equation}
H_{\rm m} = H_0 + H_1 + \mathcal{O}\left(\theta_0^2\right) \text{,  with  } H_1 \equiv
V(r) \begin{pmatrix} 
    0 & -\theta_0 \\
    -\theta_0  & 0
\end{pmatrix}.
\end{equation}

Our final state free particle of momentum $\vec{k}_f$ is characterized by the plane wave $\psi_s = e^{i \vec{k}_f . \vec{r}}$.
The mostly-active states inside the spherically symmetric bottle are 
$\psi_a(r,\Omega) = \psi_{l}(r, k) Y_{lm}(\Omega)$,
with $Y_{l m}(\Omega)$ being the spherical harmonics of the angular momentum state $\ket{l, m}$, and $\psi_{l}(r,k)$ is the radial wavefunction for an initial momentum $k = |\vec{k}|$.
We assume that the initial kinetic energy, $E = k^2/2m_n$, is small compared to the potential energy, $V_0$.
Choosing the initial to be in the $s$-wave, $l = m = 0$, the centrifugal potential vanishes, and we can express the radial dependence as
\begin{equation}
    \chi(r,k) = r \psi_{l}(r,k) = \sqrt{\frac{2}{R}} \times 
    \begin{cases}
        \sin (k r), &  r < R,\\
        A e^{- \lambda (k - R)}, & r > R,\\
    \end{cases}
\end{equation}
where $\lambda^2 = 2 m_n (V_0 - E)$ is the decay constant and $A$ some normalization factor.
The continuity condition for the wavefunction and its derivatives yields $|A| = \sqrt{E/V_0}$, which is smaller than unity for slow neutrons.
Note also that $k R \ll 1$ for UCNs.

Fermi's golden rule then gives us the transition rate,
\begin{equation}
    \Gamma = 2\pi |\bra{\psi_s} H_1 \ket{\psi_a}|^2 \delta\left( \frac{k^2 - k_f^2}{2 m_n} -  \delta m \right) \frac{\dd^3 k_f}{(2 \pi)^3},
\end{equation}
where the delta function ensures energy conservation.
The amplitude of transition into an outgoing $s$-wave can be explicitly calculated.
For $r>R$, the transition rate is finite and given by
\begin{equation}\label{eq:transition}
    \Gamma = \theta_0^2 t_f^{-1}  \frac{k}{k_f} \frac{\sin^2\left( k_f R + \phi \right)}{1 - \delta m/V_0}, \ \ \ \sin{\phi_f} = \frac{1}{\sqrt{1 + (\lambda/k_f)^2}}~,
\end{equation}
where $t_f = {R}/{v}$ is the inverse time of flight of the initial state with velocity $v$.
As the boundary of the potential becomes fuzzier, we can average over the radius $R$, and the oscillations will be washed out, $\sin^2(k_f R + \phi) \to \frac{1}{2}$.

The result in \Cref{eq:transition} requires some further interpretation.
Let us take, for instance, the limit of an infinite potential well, $V_0 \to \infty$ (keeping the expansion parameter $\theta V_0$ small).
In that case, the wavenumbers of the trapped states are quantized, and $k R = N \pi$, with $N$ an integer.
In that case, the transition probability in a time $t$ is obtained from:
\begin{equation}\label{eq:FinalWall}
    \Gamma \simeq 2 \theta_0^2 t_f^{-1} \frac{k}{k_f} \sin^2\left(k_f R\right) \longrightarrow P_{nn^\prime} = 2 \theta_0^2 \frac{t}{t_f} \frac{\sin^2\left(\delta m  \, t_f\sqrt{1 - \frac{2 m_n \delta m}{k^2}}\right)}{\sqrt{1 - \frac{2 m_n \, \delta m}{k^2}}}.
\end{equation}
Identifying the energy splitting induced by the magnetic field with $\mu B \to - \delta m$, $\tau$ with $2t_f$, and $L$ with $2R$, we recover Eq.~(21) of Ref.~\cite{Kerbikov:2008qs} for $\delta m \ll V_0$.
In the limit $\delta m \to 0$, for constant $\theta_0$, oscillations vanish.
On the other hand, for very large containers, $t_f$ is large, and oscillations are averaged-out.
Note that rate is only suppressed for endothermic reactions (large $\delta m >0$) when $k_f R \ll 1$, as otherwise they can be enhanced.
The factor of $k/k_f$ in the expression above is unusual, and we will discuss it below. 

\subsection{Scattering on an atom} 

Let us now consider a toy model of neutron scattering on an atom, represented as a spherically symmetric potential,
\begin{equation}
    V = \theta V_0 \Theta(R_0 - r),
\end{equation}
where $R_0$ quantifies the range of the interaction.
The cross section can be calculated in pertubation theory to give 
\begin{equation}\label{eq:FinalAtom}
    \sigma = \frac{\theta_0^2 m_n^2V_0^2}{\pi}\frac{k_f}{k}  \left|\int_0^{R_0} \dd r \frac{\sin(kr)\sin(k_f r)}{k \, k_f} \right|^2 =     
\frac{\theta_0^2 m_n^2V_0^2}{\pi}\frac{k_f}{k} \left(\frac{k \cos{(k R_0)} \sin{(k_f R_0)} - k_f\cos{(k_f R_0)} \sin{(k R_0)}}{2 k \,k_f \, m_n \, \delta m}\right)^2.
\end{equation}
For a small-range interaction, with $k R_0, k_fR_0 \ll 1$, we recover the geometric limit, 
\begin{equation}
    \sigma=\frac{\theta_0^2 R_0^2}{\pi} \sqrt{1 - \frac{2 m_n \, \delta m}{k^2}}\left(\frac{m_n R_0^2 V_0}{3}\right)^2.
\end{equation}
We can then conclude that short-range interactions with small objects are indeed phase space-suppressed for endothermic (large $\delta m$) transitions.
Comparing with the expression in \Cref{eq:FinalWall}, we see that under the similar limit $k_f R \ll 1$, when the neutron wavelength is much larger than the bottle size, we obtain a similar suppression (as we are also justified in expanding the sine function).
In the general case, however, the proportionality to $k_f/k$ is not guaranteed, neither for neutron scattering on bottle walls as in \Cref{eq:FinalWall} nor for scattering on an atom as in \Cref{eq:FinalAtom}.

\bibliographystyle{apsrev4-1}
\bibliography{main}{}
\end{document}